\documentclass[a4paper]{article}
\usepackage{comment}
\usepackage{xcolor}
\usepackage{INTERSPEECH2021}
\usepackage{mathtools}
\usepackage[bookmarks=false]{hyperref}
\usepackage{xurl}

\title{FRILL: A Non-Semantic Speech Embedding for Mobile Devices}
\name{Jacob Peplinski$^1$, Joel Shor$^2$, Sachin Joglekar$^3$, Jake Garrison$^3$, Shwetak Patel$^{1,3}$}

\address{
  $^1$University of Washington, Seattle, USA\\
  $^2$Google, Japan\\
  $^3$Google, USA}
\email{jpeplins@uw.edu}

\begin{document}
\maketitle

\begin{abstract}
Learned speech representations can drastically improve performance on tasks with limited labeled data. However, due to their size and complexity, learned representations have limited utility in mobile settings where run-time performance can be a significant bottleneck. In this work, we propose a class of lightweight non-semantic speech embedding models that run efficiently on mobile devices based on the recently proposed TRILL speech embedding. We combine novel architectural modifications with existing speed-up techniques to create embedding models that are fast enough to run in real-time on a mobile device and exhibit minimal performance degradation on a benchmark of non-semantic speech tasks. One such model (FRILL) is 32x faster on a Pixel 1 smartphone and 40\% the size of TRILL, with an average decrease in accuracy of only 2\%. To our knowledge, FRILL is the highest-quality non-semantic embedding designed for use on mobile devices. Furthermore, we demonstrate that these representations are useful for mobile health tasks such as non-speech human sounds detection and face-masked speech detection. Our models\footnote[1]{\url{https://tfhub.dev/s?q=nonsemantic-speech-benchmark\%2Ffrill}} and code\footnote[2]{\url{https://github.com/google-research/google-research/tree/master/non\_semantic\_speech\_benchmark}} are publicly available.
\end{abstract}


\section{Introduction}
Representation learning is a powerful tool for leveraging large collections of unlabeled data to learn better supervised models when labels are scarce \cite{wav2vec2, unsupspeech, mockingjay}. Shor et al. recently proposed the "NOn-Semantic Speech Benchmark" (NOSS) for evaluating the quality of universal speech representations \cite{trill}. NOSS includes a diverse set tasks such as emotion recognition \cite{crema, savee}, speaker identification \cite{voxceleb}, language identification \cite{voxforge}, and keyword detection \cite{speechcommands}, and is designed to encourage the development of non-semantic speech embeddings. Shor \cite{trill} also proposes a baseline representation named \textit{TRIpLet-Loss Network} (TRILL), which performs best over all NOSS benchmark tasks and achieves state-of-the-art results in some.

Many of the tasks in the NOSS benchmark, such as keyword detection and speaker identification, have natural mobile computing applications (e.g. verifying a user and triggering a voice assistant). On a mobile device, a non-semantic speech embedding could be used as input features for several real-time audio detection tasks, considerably reducing the cost of running models simultaneously. Such an embedding could enable mobile devices to listen for additional events such as non-speech health sounds (e.g. coughing, sneezing) with minimal impact on battery performance. This is desirable as real-time analysis of mobile audio streams has shown to be useful for tracking respiratory symptoms \cite{whosecough, al2020flusense, listen2cough}.

However, TRILL is based on a modified version of ResNet50 \cite{hershey2017cnn}, which is expensive to compute on mobile devices. The TRILL authors addressed this by distilling TRILL to a student model comprised of a truncated MobileNet architecture \cite{mobilenet} and two large dense layers (TRILL-Distilled), which showed minimal performance degradation on most NOSS tasks. Due to the size of its final dense layers, TRILL-Distilled contains over 26M parameters, which is still too large to run in real-time on many devices.

In this work, we address this gap by creating non-semantic speech embeddings that are fast and small enough to run in real-time on mobile devices. To do this, we use knowledge distillation \cite{distillation} to train efficient student models based on \textit{MobileNetV3} \cite{mnetv3} to mimic the TRILL representation. We apply a combination of novel architectural modifications and existing speed-up techniques such as low-rank matrix approximation \cite{svd, lowrank} and weight quantization \cite{quant} to further optimize student embeddings. Finally, in addition to the NOSS benchmark, we assess the quality of our embeddings on two privacy-sensitive, health-sensing tasks: human sounds classification \cite{esc50} and face-mask speech detection \cite{compare2020}. In summary, our main contributions are:

\begin{enumerate}
    \item Create a class of non-semantic embedding models that are fast enough to run in real-time on a mobile device. One model, which we name FRILL, is 32x faster and 40\% the size of TRILL, with an average decrease in accuracy of only 2\% over 7 diverse datasets. FRILL is 2.5x faster and 35\% the size of TRILL-distilled.
    \item Evaluate the impact of performance optimization techniques like quantization-aware training, model compression, and architecture reductions on the latency, accuracy, and size of our embedding models.
    \item Benchmark our on-device representations on two mobile-health tasks: a public dataset of human sounds, and detecting face-masked speech.
\end{enumerate}

\section{Student Model Architecture}
Our student models map log Mel-spectrograms to an embedding vector and are trained to mimic the TRILL representation. The student model architecture consists of two components: a \textit{MobileNetV3} variant (\textit{see \ref{mnet_size}}) followed by a fully-connected \textit{bottleneck} layer. \textit{MobileNetV3} extracts rich information from inputted log Mel-spectrograms and the bottleneck layer ensures a fixed embedding size. To explore the tradeoff between the performance and latency of our student models, we propose and vary a set of hyperparameters which we describe below.

\subsection{MobileNet Size} \label{mnet_size}
\textit{MobileNetV3} was officially released in two sizes: Small and Large. The small variant is targeted toward resource-constrained applications and contains fewer inverted residual blocks and convolutional channels. In addition to the official sizes, we propose a truncated version of \textit{MobileNetV3Small} which we name \textit{MobileNetV3Tiny}. It features the following modifications:
\begin{itemize}
  \item We remove two of the eleven inverted residual blocks (blocks 6 and 11) from \textit{MobileNetV3Small}. These blocks were chosen because they are duplicates of the preceding block.
  \item We reduce the number of channels in the final convolutional layer from 1024 to 512.
\end{itemize}
See Table 2 in \cite{mnetv3} for a full specification of \textit{MobileNetV3Small}.

\subsection{MobileNet Width}
MobileNet architectures feature a width multiplier $\alpha$ which modifies the number of channels in the convolutional layers within each inverted residual block. This hyperparameter is commonly used to exchange model latency for performance and has been included in other efficient neural network architectures \cite{shufflenet}. The values of $\alpha$ we explore are shown in Table \ref{table:hparams}.

\subsection{Global Average Pooling}
\textit{MobileNetV3} produces a set of two-dimensional feature maps at its output. When global average pooling (GAP) is disabled, these features maps are flattened, concatenated, and passed to the bottleneck layer to produce an embedding. This concatenated vector is large, resulting in a sizeable kernel in the bottleneck layer. GAP can be used to reduce the size of the bottleneck layer kernel by taking the global average of all "pixels" in each output feature map, thus reducing the size of the bottleneck input. Our intuition for doing this is that GAP discards temporal information within an input audio window, which is less important for learning a non-semantic speech representation due to the fact non-lexical aspects of the speech signal (e.g. emotion, speaker identity) are more stable in time compared to lexical information.

\subsection{Bottleneck Layer Compression} 
A significant portion of our student model weights are located in the kernel matrix of the bottleneck layer. To reduce the footprint of this layer, we apply a compression operator based on Singular Value Decomposition (SVD) that learns a low-rank approximation of the bottleneck weight matrix $W$\footnote[3]{\url{https://github.com/google-research/google-research/tree/master/graph_compression}}. Similar to \cite{lowrank}, we aim to learn the low-rank approximates during training as opposed to post-training. Formally, this operator uses SVD to create matrices $U$ and $V$ such that the Frobenius norm of $W-UV^\intercal$ is minimized. The compressed kernel replaces a matrix of $m \times n$ weights with $k(m + n)$ weights, where $k$ is a hyperparameter that specifies the inner dimension of $U$ and $V$, which we fix at $k=100$ for this study. A convex combination of original and compressed kernels is used during training to produce the following layer output:

\begin{equation}
     y = x(\lambda W + (1-\lambda)UV) + b
\end{equation}

where $b$ is the bias vector in the bottleneck layer, $x$ is the input vector, and $\lambda$ is a scalar that is set to one at the beginning of training and linearly decreases to zero over the first ten training epochs. Varying $\lambda$ helps the optimizer transition to learning the weights of the compressed matrices. At inference time, $\lambda$ is explicitly set to 0 and the original kernel is discarded.

\subsection{Bottleneck Layer Quantization}
Quantization aims to reduce model footprint and latency by reducing the numerical precision of model weights. Instead of using post-training quantization which can cause performance degradation, we use \textit{Quantization-Aware Training} (QAT), a procedure which gradually quantizes model weights during training. We use the Tensorflow implementation of QAT to quantize the bottleneck layer kernel from 32-bit floating point to 8-bits, which is based on the quantization scheme described in \cite{quant}.

\begin{table}[t]
\setlength{\tabcolsep}{3pt}
\small
\begin{centering}
\caption{Model hyperparameters for reducing size and latency.}
\label{table:hparams}
\begin{tabular}{ccc}
\toprule[2pt]
Name & Description & Values \\
\toprule[2pt]
MV3Size  & MobileNetV3 size & tiny, small, large \\
\midrule
MV3Width & MobileNet width  & \begin{tabular}{@{}c@{}}0.5, 0.75, 1.0, \\ 1.25, 1.5, 2.0\end{tabular} \\
\midrule
GAP      &   Global average pooling & yes, no \\
\midrule
Compression &   Bottleneck compression & yes, no \\
\midrule
QAT &   Quantization-aware training & yes, no \\
\bottomrule
\end{tabular}
\end{centering}
\vspace{-4ex}
\end{table}

\section{Experiments}

We conduct a study to determine the effect of each hyperparameter in Table \ref{table:hparams} on the representation quality, latency, and size of our student embedding models. For each of 144 combinations of hyperparameters, we distill the TRILL embedding to a student network, benchmark the student embedding by training simple classifiers to solve NOSS tasks and health tasks using embeddings as input features, and measure inference latency on a Pixel 1 smartphone. The distillation dataset, student network training procedure, NOSS benchmarking, and latency benchmarking procedures are described in the following sections.

\subsection{Distillation Dataset}
To build our dataset for distillation, we randomly sample a 0.96-second audio context from each Audioset \cite{audioset} speech clip and compute a log-magnitude Mel spectrogram using a Short-Time Fourier Transform (STFT) window size and window stride of 25ms and 10ms respectively. We compute 64 Mel bins. Using each spectrogram, we compute the \textit{layer19} output of the TRILL model, as was done in \cite{trill}. Each \{log Mel spectrogram, \textit{layer19}\} pair is stored as a single observation for distillation training. 
Because some YouTube videos were unavailable at the time of this study, we were only able to procure 902,523 clips, which accounts for 89.2\% of the published speech subset.

\subsection{Student Model Training}
\begin{figure}[t]
\begin{minipage}[t]{1\linewidth}
  \centering
  \vspace{-.3cm}
  \centerline{\includegraphics[width=8.0cm]{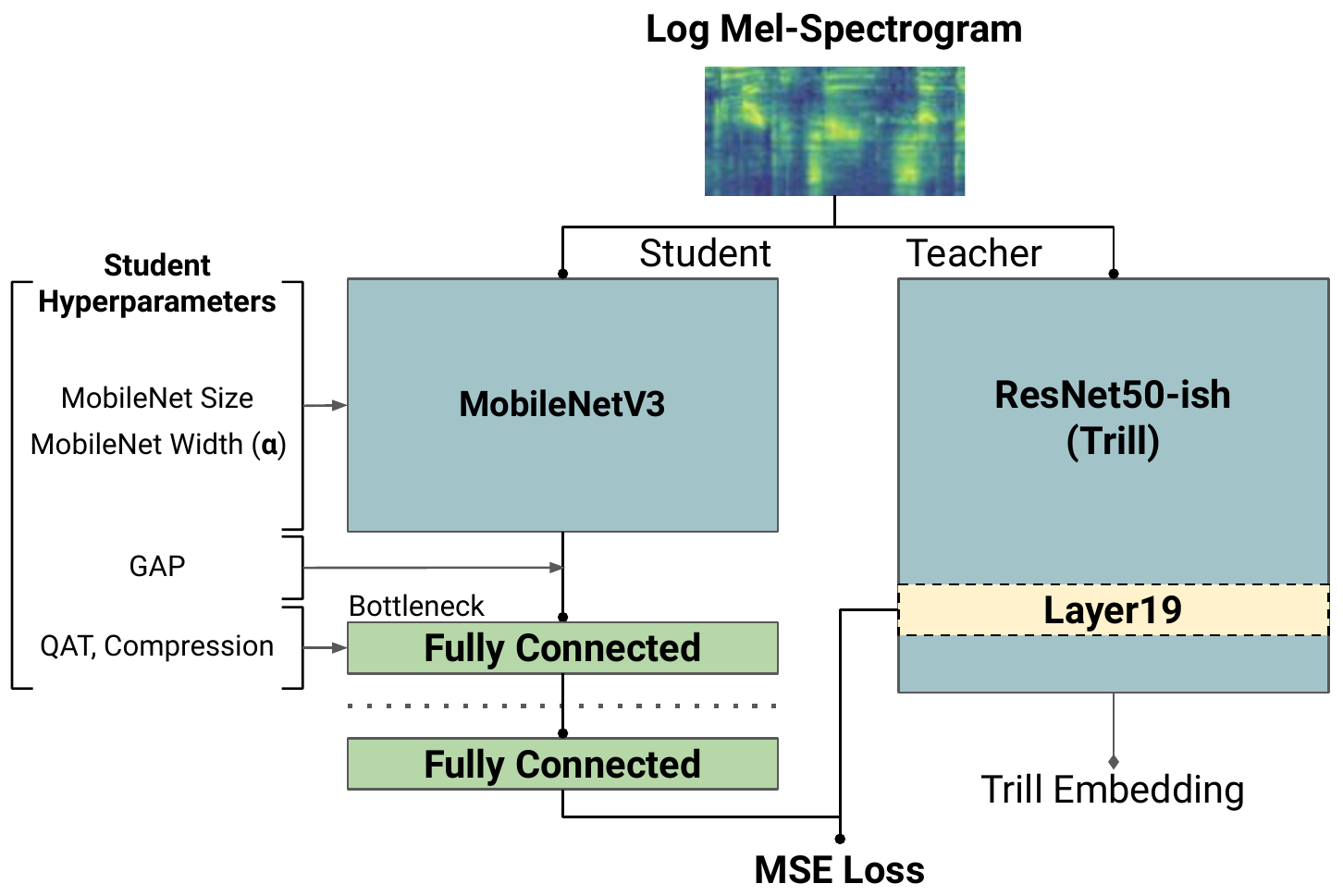}}
\caption{Knowledge distillation for non-semantic speech embeddings. The dashed line shows the student model's output.}
\label{fig:model-diagram}
\end{minipage}
\setlength{\belowcaptionskip}{-10pt}
\end{figure}

Student models are trained to map input spectrograms to the \textit{layer19} representation produced by TRILL. Because the \textit{layer19} vector is much larger (12288d) than our student embeddings (2048d), we append an equal-length fully-connected layer to the output of the student model. This layer allows us to take a mean-squared-error loss against \textit{layer19}. A diagram of our training setup is shown in Figure \ref{fig:model-diagram}. To train student models, we use a batch size of 128 and an initial learning rate of 1e-4 with the Adam optimizer \cite{adam}. We use an exponential learning rate schedule, with learning rates decreasing by a factor of 0.95 every 5,000 training steps. Each model trains for 50 epochs, or approximately 350,000 training steps.

\subsection{NOSS Benchmark Analysis}
To evaluate the quality of our student embeddings, we train a set of simple classifiers using embeddings as input features to solve each classification task in the NOSS benchmark. As detailed in \cite{trill}, for each dataset in NOSS, we train a logistic regression, random forest, and linear discriminant analysis classifier using the SciKit-Learn library \cite{sklearn}. Embeddings for each utterance are averaged in time to produce a single feature vector. For tasks that contain multiple observations per speaker (\textit{SpeechCommands}, \textit{CREMA-D}, \textit{SAVEE}) we also train a set of classifiers using L$^{2}$ speaker normalization, as in \cite{trill}. We report the best test accuracy across combinations of downstream classifiers and normalization techniques. Accuracies on DementiaBank \cite{dementiabank}, one of the datasets included in the original NOSS benchmark, were all within 1\% of each other, so we excluded it from our analysis in Section \ref{results}.

\subsection{Mobile Health-Sensing Tasks}
In addition to tasks in the NOSS benchmark, we evaluate Trill, Trill-Distilled, and each of our student models on a human sounds classification task and a face-mask speech detection task. The human sounds task is derived from the ESC-50 dataset \cite{esc50}, which contains 5-second sound clips from 50 classes. The human sounds subset of this dataset constitutes 10 of the 50 classes and includes labels such as 'coughing', 'sneezing', and 'breathing'. Similar to NOSS, we train a set simple classifiers using input features from each student model and report test accuracy on the best model. We use the first four published folds of ESC-50 for training, and the fifth for testing.

The objective of the mask speech task is to detect whether 1-second speech clips are from masked or unmasked speakers \cite{compare2020}. The dataset contains around 19,000 masked and 18,000 unmasked speech examples. Although the test set labels were not available at the time of this publication, and the baseline publication evaluates models on the unweighted average recall instead of accuracy, we track our models' performance here as an indicator of their suitability for mobile health tasks.

\subsection{Run-time Analysis}
\label{sec:runtime}
The TensorFlow Lite (TFLite) framework enables execution of machine learning models on mobile and edge devices. To measure the run-time performance of our student embeddings in their intended environment, we convert each model to TFLite's \textit{flatbuffer} file format for 32-bit floating-point execution and benchmark inference latency (single-threaded, CPU execution) on the Pixel 1 smartphone. We also verified conversion to the \textit{flatbuffer} format does not effect the quality of our representations. Latency measurements for TRILL and TRILL-Distilled have also been recorded for reference.

\section{Results} \label{results}

\begin{table*}[t]
\setlength{\tabcolsep}{5pt}
\small
\begin{centering}
\caption{Test Performance on the NOSS Benchmark and Mobile Health Tasks. Sample of model performances and latencies on the quality/latency tradeoff curve. *Masked Speech test set labels are not available at this time, so we report accuracy on the eval set.}
\label{tab:best_models}
\begin{tabular}{@{} c|cccccccccc @{}}
\toprule[2pt]
Model & Voxceleb1 & Voxforge & \begin{tabular}{@{}c@{}}Speech \\ Commands\end{tabular}   &  CREMA-D & SAVEE & \begin{tabular}{@{}c@{}}Masked$^*$ \\ Speech\end{tabular}  & \begin{tabular}{@{}c@{}}ESC-50 \\ HS\end{tabular} & \begin{tabular}{@{}c@{}}Size \\ (MB)\end{tabular} & \begin{tabular}{@{}c@{}}Latency \\ (ms)\end{tabular}\\
\toprule[2pt]
TRILL           & 48.5 & 84.5 & 81.9 & 66.2 & 70.0 & 66.0 & 86.4 & 98.1  & 275.3 \\
TRILL-Dist      & 47.4 & 80.0 & 80.2 & 70.2 & 70.0 & 67.2 & 87.9 & 107.1 &  22.5 \\
\midrule
Small\_2.0\_GAP (FRILL) & 44.5 & 76.9 & 79.7 & 70.9 & 67.5 & 65.7 & 86.4 & 38.5 & 8.5 \\  
Small\_0.5\_QAT       & 37.0 & 75.3 & 76.6 & 67.0 & 67.5 & 63.4 & 77.3 & 12.7 & 3.0 \\  
Tiny\_0.5\_Comp\_GAP & 29.2 & 68.0 & 57.8 & 60.8 & 59.2 & 61.6 & 78.8 & 2.3  & 0.9 \\  
\bottomrule
\end{tabular}
\end{centering}
\label{table:noss}
\end{table*}


\begin{figure}[t]
  \centering
  \vspace{-.3cm}
  \centerline{\includegraphics[width=8.0cm]{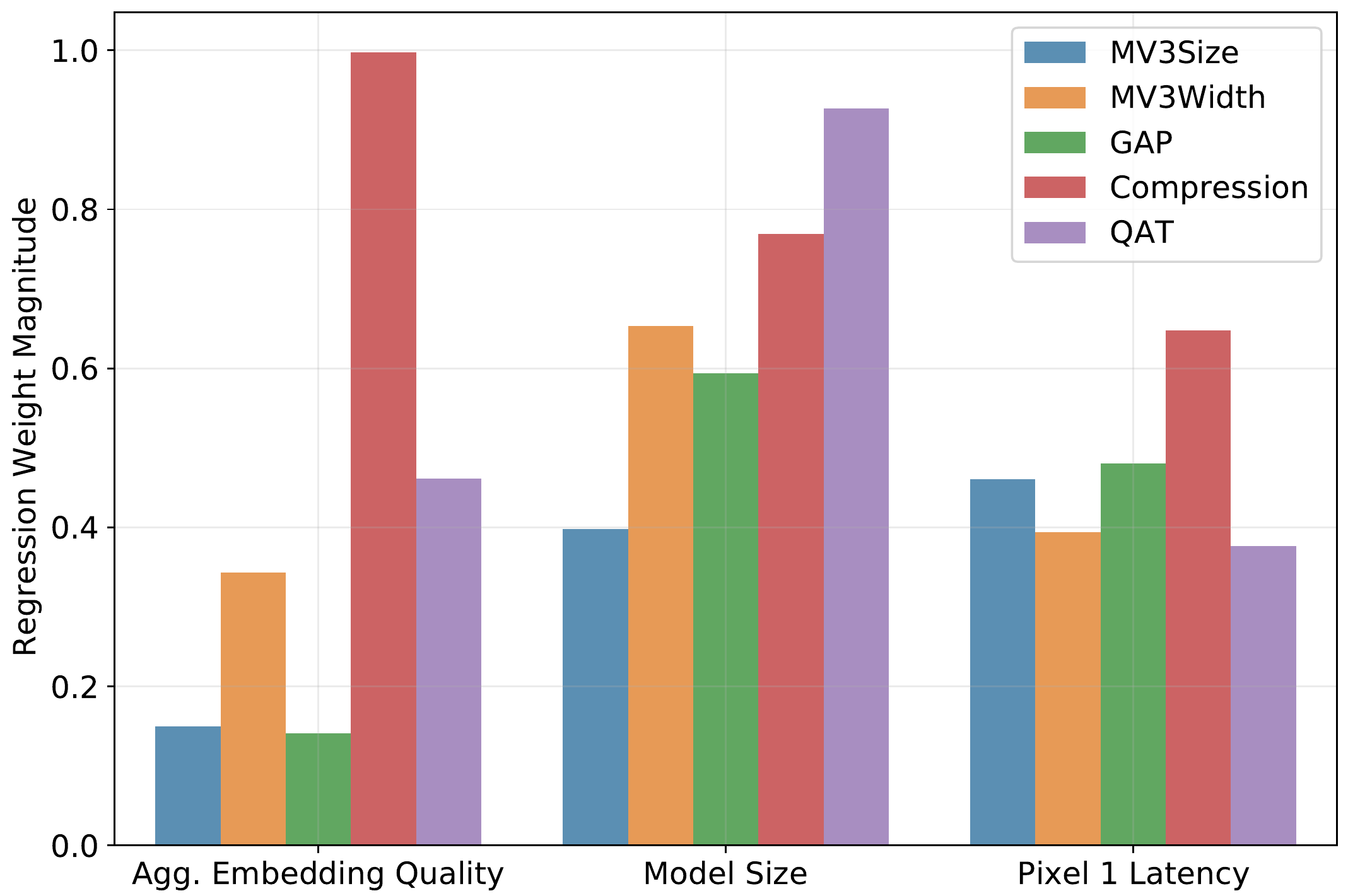}}
\caption{Linear regression weight magnitudes for predicting model quality, latency, and size. The weights indicate the expected impact of changing the input hyperparameter. A higher weight magnitude indicates a greater expected impact.}
\label{fig:regression}
\end{figure}



Because student embeddings are evaluated on 7 datasets, it is difficult to naturally rank models based on their "quality". Thus, we devise an \textit{Aggregate Embedding Quality} score by computing the performance difference between a student model and TRILL for each task, and averaging across tasks: 
\begin{equation}
\text{Aggregate Embedding Quality}_m = \frac{1}{|D|}\sum_{d}(A_{md} - T_d)
\end{equation}
where $m$ indicates the student model, $d$ indicates the dataset, and $T_d$ is the accuracy of TRILL on dataset $d \in D$. This score tells us the average deviation from TRILL's performance across all NOSS tasks and mobile health tasks.

To understand the impact each hyperparameter in Table \ref{table:hparams} has on our student models, we perform a multivariate linear regression to model aggregate quality, latency, and size using model hyperparameters as predictors. We standardize each regression target in order to produce regression weights on the same order of magnitude while preserving relative importance. The magnitude of regression weights are shown in Figure \ref{fig:regression}.

\begin{figure}[t]
  \centering
  \vspace{-.3cm}
  \centerline{\includegraphics[width=7.9cm]{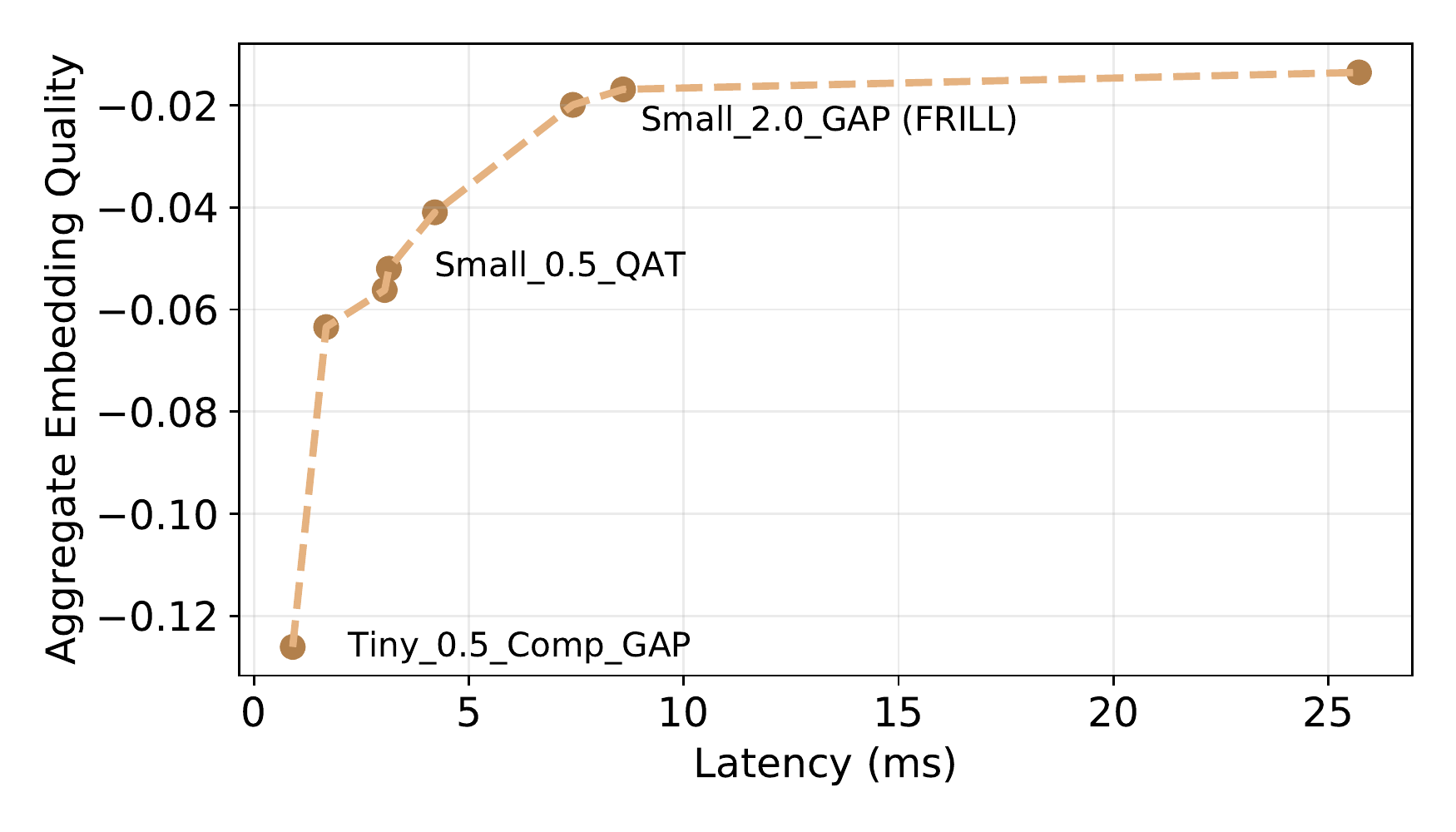}}
\caption{Embedding quality and latency tradeoff. x-axis is the inference latency, y-axis is the difference in accuracy from TRILL's performance, averaged across benchmark datasets.}
\label{fig:pfrontier}
\end{figure}

To illustrate the latency and quality tradeoff in our cohort of models, we produce a "quality" frontier plot. For all latency measurements $l$ in our study, we pick the model with the best aggregate embedding quality with a latency less than or equal to $l$. This frontier, shown in Figure \ref{fig:pfrontier}, features 8 student models of various qualities and latencies. NOSS benchmark and mobile health task accuracies are shown in Table \ref{tab:best_models} for three representative frontier models. The most performant of these representative models, which we name \textbf{FRILL} (\textit{fast TRILL}), has an aggregate embedding quality score of -0.0169, indicating an average deviation from TRILL quality of 1.69\% with respect to the datasets in this study. FRILL has an inference latency of 8.5ms on a Pixel 1 smartphone, and is only 38.5 megabytes in the TFLite file format.

\section{Discussion}
\subsection{Factors contributing to model quality and latency}
\textbf{Architecture reduction techniques have a smaller impact on performance and latency}: Reducing MobileNetV3 size via $\alpha$, by removing residual blocks, and by pooling early in the network had a smaller effect than QAT and bottleneck compression (Figure \ref{fig:regression}). This suggests that the TRILL-distilled mobilenet part of the architecture was overparameterized compared to the representation quality possible by the bottleneck.

\textbf{QAT reduces model size the most and latency the least}: QAT reduces overall model size the most and pixel 1 latency the least (Figure \ref{fig:regression}). It decreases embedding quality by only half as much as compression, and is present in 1/8 of the best models.

\textbf{Bottleneck compression reduces embedding performance the most}: This suggests that TRILL-distilled's last, bottleneck layer is one of the most performance-sensitive parts of the model. More sophisticated methods than low-rank approximation might be necessary. Compression is only present in 1/8 of the frontier models.

\subsection{Quality / latency tradeoff}

After eliminating models with better and faster alternatives, we are left with 8 "frontier" models (Figure \ref{fig:pfrontier}). The fastest model runs at 0.9 ms, which is 300x faster than TRILL and ~25x faster than TRILL-distilled. FRILL runs at 8.5 ms, which is about 32x faster than TRILL 2.5x faster than TRILL-distilled. FRILL is also roughly 40\% the size of TRILL and TRILL-distilled.

The curve is steep on both sides of our frontier. This means that with minimal latency costs we can achieve much better performance on one end, and vice versa on the other. This supports our choice of experiment hyperparameters. Though there is a frontier model with an aggregate embedding quality higher than FRILL, it comes at the cost of a significant bump in latency.

\subsection{Limitations and Future Work}

In this paper, we focus on making the TRILL embedding smaller and faster for mobile applications. Some of our optimization procedures (architecture reduction and bottleneck compression) take advantage of TRILL-specific model features, so it is unclear the extent to which these results generalize to other representations. In the future, we would like to make other representation models more lightweight.

Our latency numbers were generated using the Pixel 1 smartphone. While this is a common phone that is resource-constrained compared to newer models, we do not know the extend to which our results generalize to even more resource-constrained environments. Future work can explore benchmarking these smaller models in more constrained environments. The smallest, fastest models might be suitable for use on smart watches or smart home devices.

Finally, having a non-semantic speech embedding in a mobile setting unlocks many privacy-sensitive applications. Future work will include benchmarking on more tasks in this category. Shor \cite{trill} demonstrated that non-semantic embedding can be fine-tuned for improved performance, and future work includes testing latency and performance for on-device training.

\section{Conclusions}
In this work, we proposed an efficient non-semantic speech embedding model trained via knowledge distillation that is fast enough to be run in real-time on a mobile device. We explore latency and size reduction techniques, and quantify their impact on model quality. We then characterize the performance / latency tradeoff curve for the 144 models we trained, and report size, latency, and performance numbers for representative models. We identify one model, FRILL, which exhibits a 32x inference speedup and 60\% size reduction, with an average decrease in accuracy of less than 2\% over 7 different datasets, as compared to the original TRILL model. FRILL is 2.5x faster and 35\% the size of TRILL-distilled. We also demonstrate the effectiveness of our embeddings on two new mobile health tasks. These new tasks in particular benefit from the on-device nature of our embeddings, since performing computations locally can improve both the privacy and latency of resulting models.

\bibliographystyle{IEEEtran}
\bibliography{foo}
\end{document}